\documentclass[pra,aps,twocolumn]{revtex4}
\usepackage[usenames,dvipsnames]{color}
\usepackage{graphicx}
\usepackage{dcolumn}
\usepackage{bm}
\usepackage{amsmath}
\usepackage{gensymb}
\usepackage{amsfonts}
\usepackage{psfrag}
\usepackage[bookmarks,colorlinks=false]{hyperref}

\renewcommand{\vec}[1]{\mathbf{#1}}
\newcommand{\Eq}[1]{Eq. (\ref{#1})}

\begin{document}
\title{Theoretical Investigation of Phonon Polaritons in SiC Micropillar Resonators}

\author{Christopher R. Gubbin$^1$}
\author{Stefan A. Maier$^2$}
\author{Simone \surname{De Liberato}$^1$}

\affiliation{$^1$School of Physics and Astronomy, University of Southampton, Southampton, SO17 1BJ, United Kingdom}
\affiliation{$^2$Department of Physics, Blackett Laboratory, Imperial College London, London SW7 2AZ United Kingdom}

\begin{abstract}
Of late there has been a surge of interest in localised phonon polariton resonators which allow for sub-diffraction confinement of light in the mid-infrared spectral region by coupling to optical phonons at the surface of polar dielectrics. Resonators are generally etched on deep substrates which support propagative surface phonon polariton resonances. Recent experimental work has shown that understanding the coupling between localised and propagative surface phonon polaritons in these systems is vital to correctly describe the system resonances. In this paper we comprehensively investigate resonators composed of  arrays of cylindrical SiC resonators on SiC substrates. Our bottom-up approach, starting from the resonances of single, free standing cylinders and isolated substrates, and 
exploiting both numerical and analytical techniques, allows us to develop a consistent understanding of the parameter space of those resonators, putting on firmer ground this blossoming technology.
\end{abstract}

\maketitle
Sub-diffraction confinement of light is a necessity for miniaturisation of optical devices and can be achieved by coupling photons to charged particles at interfaces over which the real part of the dielectric function changes sign. 
Predominantly free electrons, well described by a Drude-like dielectric function, are utilised \cite{MaierBook}.
As the Drude dielectric function is negative below the plasma frequency such materials provide broadband localisation of light in modes termed surface plasmon polaritons, although this comes at the cost of high loss due to electron scattering. More recently sub-diffraction confinement by coupling to crystal vibrations, in the form of optical phonons in polar dielectrics has been demonstrated \cite{Hillenbrand02}. These materials have a negative dielectric function between the longitudinal and transverse optical phonon frequencies, in a region termed the Reststrahlen band, and as the anharmonicity driven optical phonon damping occurs two orders of magnitude slower than electron damping in metals the resultant modes are comparatively long lived  \cite{Khurgin15}. These modes are termed surface phonon polaritons \cite{Borstel74} and have morphology-dependent characteristics which allow for tuning of modal frequencies and field profiles \cite{Mutschke99}. Their energies lie between the mid- and far-infrared, dependent on material choice, meaning these systems can make excellent narrowband thermal sources \cite{Greffet02,Schuller09}. Most recently, fabrication advancements have allowed for the construction of user-defined cylindrical SiC nano-resonators on SiC substrates, which exhibit quality factors in excess of the theoretical limit for plasmonic resonators, and Purcell factors 4 orders of magnitude larger than comparable plasmonic systems \cite{Caldwell13,Chen14}. Only few months ago the strong coupling between localised modes of user-defined cylindrical resonators and propagative surface phonon polaritons sustained by the substrate planar interface was demonstrated \cite{Gubbin16}, highlighting how the resonant coupling between localised and propagative modes has to be taken into account to correctly describe the optical response of such systems. Surface phonon polaritons have also been utilised for a number of other applications like sensing \cite{Hillenbrand02},  superlensing \cite{Taubner06}, near field optics \cite{Taubner04}, and enhanced energy transfer between nanoparticles \cite{Shen09}. 
Surface phonon polaritons offer great promise as a testbed for integrated, mid-infrared quantum photonics as a result of their long modal lifetimes and relative simplicity of fabrication. In order to allow for a fast development of this blossoming research field, 
 in this paper we develop, using both analytical and numerical methods, a consistent understanding of the phonon polariton resonances of periodic arrays of cylindrical SiC resonators, of the kind that have been used in recent groundbreaking works \cite{Caldwell13,Chen14,Gubbin16}. The relative simple geometry of such samples allows us to methodically explore its parameter space, gaining both a quantitative understanding of those resonators, that can directly be experimentally exploited, and also obtaining precious insights into the underlying physics, that can serve as lead to extend investigations to novel and more complex structures.
 
In a bottom-up approach we start our investigation in Sec. \ref{SingleCylinder} by considering resonances of a free standing cylindrical SiC resonator, analysing how both the resonant frequencies and a number of figures of merit change with morphology. In Sec. \ref{CylinderSubstrate} we consider the effect of a SiC substrate sitting below the cylinder, analysing how the presence of phonon polaritons on the substrate surface modify the resonant frequencies. Finally in Sec. \ref{CylinderArray} we consider an array of cylindrical resonators over a substrate, where both the folding of surface modes due to the periodic patterning and the dipolar interaction between different pillars lead to the appearance of strongly dispersive features.

\section{Free standing cylindrical SiC resonator}
\label{SingleCylinder}
\begin{figure*}
\includegraphics[width=0.7\textwidth]{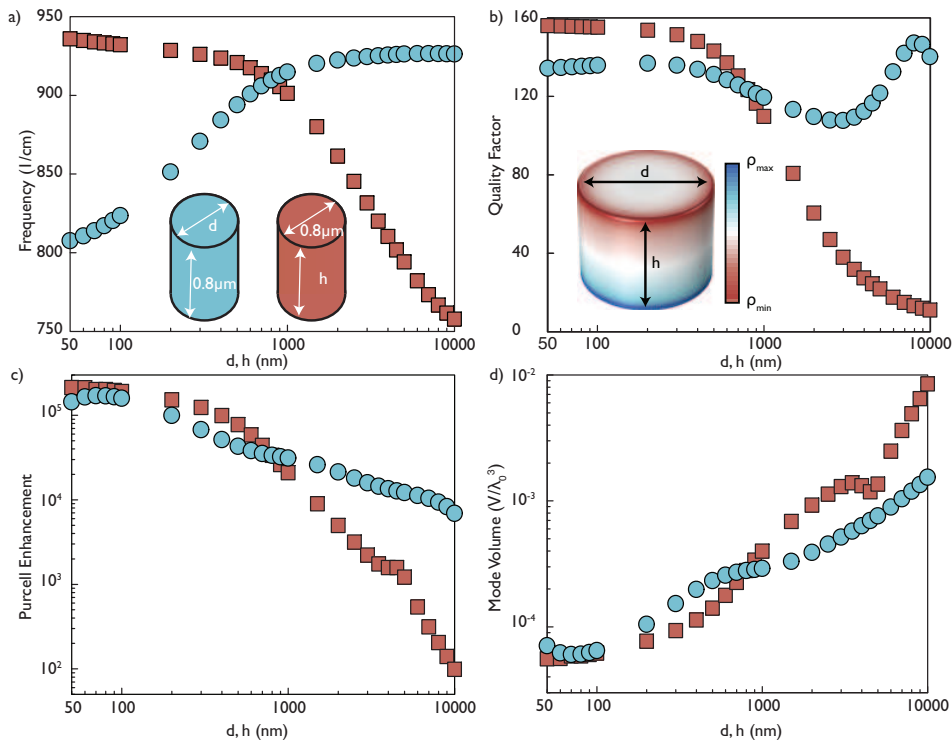}
\centering
\caption{\label{fig:Figure1}  
a) Real part of the resonant frequency of the longitudinal mode as a function of diameter at $\mathrm{h}=0.8\mu$m height (circles) and as a function of height at $\mathrm{d}=0.8\mu$m diameter (squares).
b) Quality factor of the longitudinal mode as a function of diameter at $0.8\mu$m height (circles) and as a function of height at $0.8\mu$m diameter (squares). Inset shows the surface charge distribution $\rho$ of the longitudinal mode for a cylinder of height $\mathrm{h}=0.8\mu$m and diameter $\mathrm{d}=1\mu$m. 
e)  Purcell enhancement of the longitudinal mode as a function of diameter at $0.8\mu$m height (circles) and as a function of height at $0.8\mu$m diameter (squares).
f) Mode volume of the longitudinal mode in units of the free space mode volume $\lambda_0^3$ as a function of diameter at $0.8\mu$m height (circles) and as a function of height at $0.8\mu$m diameter (squares).}
\end{figure*}
A single cylindrical resonator is characterised by two geometric parameters, its height $\mathrm{h}$ and its diameter $\mathrm{d}$. In the asymptotic limit $\mathrm{h} \to \infty$ the system is exactly described by Mie-like theories and the resonator supports a predictable series of modes characterised by polarisation and azimuthal phase dependancy \cite{Bohren}. For finite $\mathrm{h}$ Fabry-P{\'e}rot modes are supported along the length of the cylinder \cite{Ditlbacher2005}. We are here interested in the optical response of cylindrical SiC resonators with deeply subwavelength $\mathrm{h}$ and $\mathrm{d}$. This system has no closed form analytical solution so we simulate numerically by finite element methods, using the RF Module of COMSOL Multiphysics.

The SiC resonators are non-conservative systems with complex modal frequencies $\tilde{\omega}_{\mathrm{m}}$ representing loss through energy leakage from the system. Even if intrinsic material losses are neglected, this leads to a non-Hermitian time-evolution operator for the isolated cylinder \cite{Changbook,Leung1994}. The modes of the system must thus be described in the formalism of quasi-normal modes \cite{Sauvan2013}. This is especially important when calculating the mode volume through the usual definition based on the systems electromagnetic energy
\begin{equation}
\mathrm{V} = \frac{\int \epsilon\left(\vec{r}\right) \left|\vec{E}\left(\vec{r}\right)\right|^2 \mathrm{d}^3 \vec{r}}{2 \epsilon_0 \mathrm{n}^2 \left|\vec{E}\left(\vec{r}_{\mathrm{max}}\right)\right|^2} ,
\label{eq:ModeVolStandard}
\end{equation}
where $\epsilon_0$ is the vacuum permittivity, $\mathrm{n}$ the environment refractive index, $\epsilon\left(\vec{r}\right)$ is the permittivity of the resonator and $\vec{E}\left(\vec{r}\right)$ is the electric field of the mode with peak value  $\vec{E}\left(\vec{r}_{\mathrm{max}}\right)$. In systems possessing a complex modal frequency $\tilde{\omega}_\mathrm{m}$ the fields diverge as $\left|\vec{r}\right|\to\infty$ resulting in divergence of the integral \cite{Kristensen13, Koenderink2010}. In addition the Kramers-Kronig consistency of the dielectric function means loss necessarily results in a dispersive dielectric function, which leads to an alteration of the integral to account for energy in the matter \cite{Ruppin2002}. These problems are solved by explicit calculation of the system quasi-normal modes as described in Appendix A.

For excitations parallel to the cylinder axis the lowest lying mode has the surface charge distribution shown in the inset of Fig.~\ref{fig:Figure1}b. It corresponds to the fundamental Fabry-P{\'e}rot resonance of the TM$_0$ mode of the cylinder. This mode was first predicted for long silver nanowires by Takahara {\it et al.} \cite{Takahara1997} and corresponds in our case to the fundamental longitudinal dipolar resonance of the cylinder. This has been discussed extensively for SiC cylinders on substrate by Caldwell {\it et al.} where it is termed the monopolar mode \cite{Caldwell13,Chen14}. We investigated the resonant frequency of this mode over the 2-dimensional parameter space, with results shown in Fig.~\ref{fig:Figure1}a. The cylinder height is varied at a constant diameter $\mathrm{d} = 0.8\mu$m,  and the diameter is varied at a constant height $\mathrm{h} = 0.8\mu$m, as schematically shown in the inset of Fig.~\ref{fig:Figure1}d. In the small diameter limit the resonant frequency tends to that of an infinite wire given by $\epsilon\left(\omega\right) \to - \infty$ at $\omega_{\mathrm{TO}}$. As the cylinder diameter is increased the resonance tends toward an asymptote at $\approx 934/$cm, slightly blue shifted from the Fr{\"o}lich resonant frequency The quality factor over this range as shown in Fig.~\ref{fig:Figure1}b is fairly flat as the length scale of the mode is unchanged. In the limit of vanishing height the longitudinal resonance should lie at the longitudinal optical phonon frequency where $\epsilon\left(\omega\right) = 0$. Over the range shown in Fig.~\ref{fig:Figure1}a a monotonous shift away from the LO phonon frequency is observed as expected for a Fabry-P{\'e}rot resonance along the cylinder length being proportional to $1/\mathrm{h}$. At around $\mathrm{h} = 5\mu$m the mode leaves the Reststrahlen band and the character of the resonance changes from a sub-diffraction localised phonon polariton to a that of a high-index dielectric resonator. This transition is accompanied by the drop off in quality factor observed in Fig.~\ref{fig:Figure1}b.
The Purcell enhancement is shown in Fig.~\ref{fig:Figure1}c, and the procedure used to calculate it, taking care of the effect of losses, is outlined in Appendix A. The field maximum is evaluated at a point $5$nm from the cylinder edge. Mode volumes are given in Fig.~\ref{fig:Figure1}d. Smaller resonators allow for tighter confinement of the field and correspondingly larger Purcell enhancements, exceeding $10^5$ in the small resonator limit. The dip in the height scanned data at $\mathrm{h} = 4\mu$m occurs as the mode energy crosses the TO phonon energy, being evanescent in nature for smaller heights and diffraction limited for larger.\\ 
This Section has focussed on the longitudinal mode which is azimuthally invariant, meaning its azimuthal mode number is $m=0$. Modes with higher azimuthal mode numbers are of course supported with angular dependance $e^{i m \phi}$ where $\phi$ is the azimuthal angle. In Appendix B we include a similar study for the lowest order $m=1$ mode which will be referred to as the transverse dipolar mode in accordance with Caldwell {\it et al.} \cite{Caldwell13}. This mode consists of parallel dipoles excited at each cylinder end-facet. In this case Purcell enhancements exceeding $10^6$ are achievable due to tighter confinement of the mode at the cylinder vertices. We also observe higher order analogues with field inversions over the cylinder long axis. For future reference the following $m=1$ mode is a transverse quadrupole mode, analogous to the transverse dipolar mode with anti-parallel dipole alignment, leading to an additional field inversion around the cylinder centre. 

\section{Cylindrical SiC resonator on a SiC Substrate}
\label{CylinderSubstrate}
\begin{figure*}
\includegraphics[width=\textwidth]{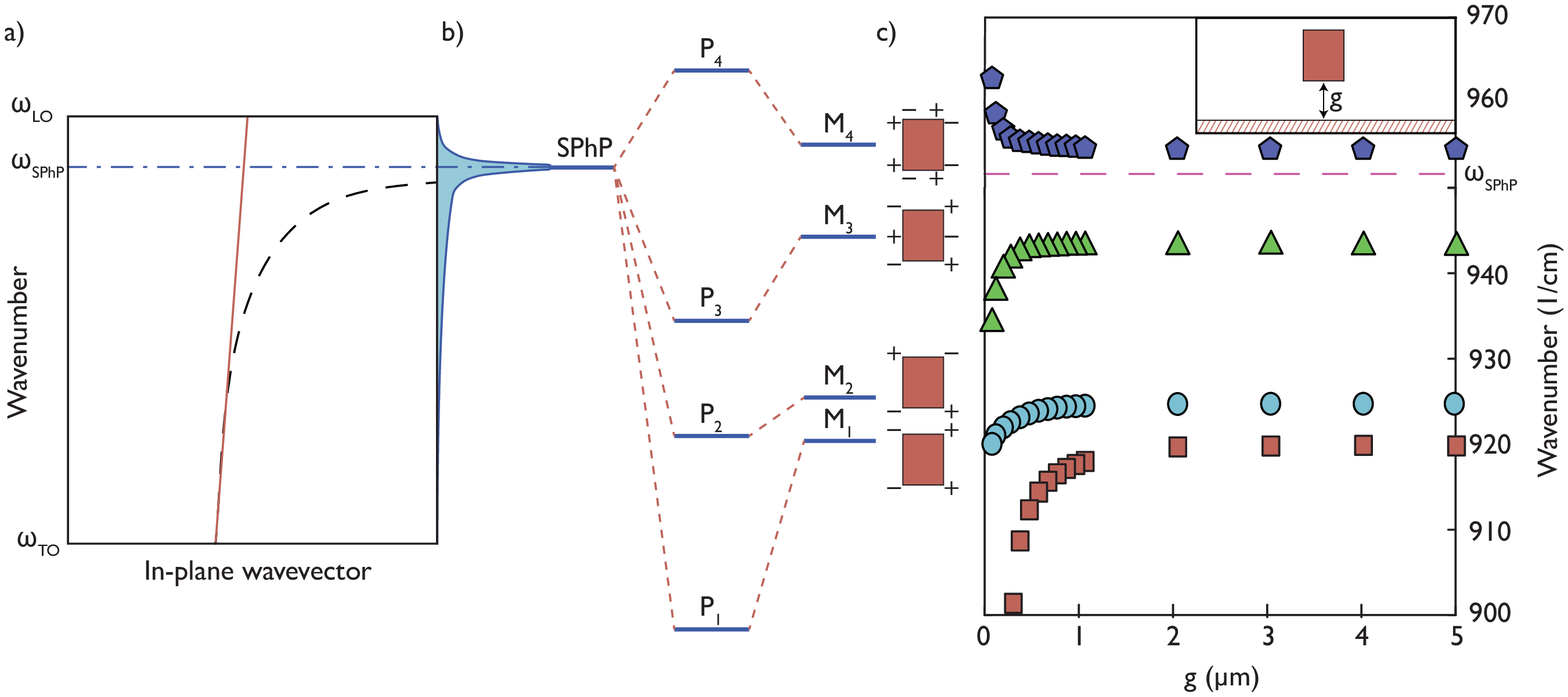}
\centering
\caption{\label{fig:Figure3} a) The dispersion of the surface phonon polariton mode supported by a vacuum/SiC interface is indicated by the black dashed line, the light cone by the red solid line and the asymptotic surface phonon polariton frequency at $\omega_{\mathrm{SPhP}}$ by a blue dot-dashed line.  The density of states is sketched on the right of the plot, peaking at $\omega_{\mathrm{SPhP}}$.
b) An illustration of the shift caused by the coupling on the resonator modes. The uncoupled modes are the surface phonon polariton (SPhP) and the lowest lying four $m=1$ modes of the resonator labelled M$_\mathrm{i}$, $\mathrm{i}=1-4$. These interact to form coupled modes labelled P$_\mathrm{i}$.
c) An illustration of the transverse mode shifts as a cylindrical resonator of height $0.8\mu$m and diameter $1 \mu$m is lowered onto a substrate.}
\end{figure*}
In the previous Section we considered the resonances of a single SiC cylinders in vacuum. The following step toward a consistent description of real resonators is to consider the effect of placing the cylinders on a SiC substrate. A 
planar, optically thick SiC substrate in vacuum supports a propagating surface phonon polariton with dispersion 	
\begin{equation}
\label{surfacedispersion}
\mathrm{k}_{\parallel} = \frac{\omega}{\mathrm{c}} \sqrt{\frac{\epsilon\left(\omega\right)}{\epsilon\left(\omega\right)+1}},
\end{equation}
where $\mathrm{k}_{\parallel}$ is the in-plane wavevector, $\epsilon\left(\omega\right)$ is the dispersive dielectric function of the substrate and $\mathrm{c}$ is the speed of light. The dispersion is plotted in Fig.~\ref{fig:Figure3}a, where is clearly visible that in the non-retarded regime the majority of the oscillator strength lies at the asymptotic frequency $\omega_{\mathrm{SPhP}}=\omega_{\mathrm{TO}}\sqrt{\frac{1+\epsilon_0}{1+\epsilon_{\infty}}} \approx951/$cm. For definiteness in the remainder of this Section we consider cylinders of diameter $\mathrm{d}=1\mu$m and height $\mathrm{h}=0.8\mu$m. For these parameters the bright transverse and longitudinal resonances lie at lower energies than the asymptote of the surface mode $\omega_{\mathrm{SPhP}}$.

While the cylinder-substrate separation is large enough, cylinder and substrate modes are good approximations for the modes of the coupled system. This ceases to be true for sub-micron distances, when the overlap of the resonator and surface modes can not be neglected. Their resulting coupling leads to repulsion between the different modes, that shift as illustrated schematically in Fig.~\ref{fig:Figure3}b. 
To illustrate this process we carry out finite element simulations of the first four $m=1$ modes of SiC cylinders discussed in Sec. I, separated from a substrate by a gap of width $g$. The relevant surface charge distribution for the different uncoupled modes of the cylinder are sketched on the left of Fig.~\ref{fig:Figure3}c, where $M_1$ and $M_2$ are the transverse dipolar and quadrupole modes described above, and $M_3$ and $M_4$ are the two following higher lying $m=1$ modes.  We plot the real parts of the resonant frequency in Fig.~\ref{fig:Figure3}c. As the resonator-substrate separation $g$ vanishes, all the modes lying below the asymptotic frequency $\omega_{\mathrm{SPhP}}$ are observed to red shift while those above blue shift. This is as expected for modes which interact with a delocalised surface mode at $\omega_{\mathrm{SPhP}}$ \cite{Nordlander2004}.\\

\section{Array of cylindrical SiC resonators on a SiC substrate}
\label{CylinderArray}

\begin{figure}
\includegraphics[width=0.35\textwidth]{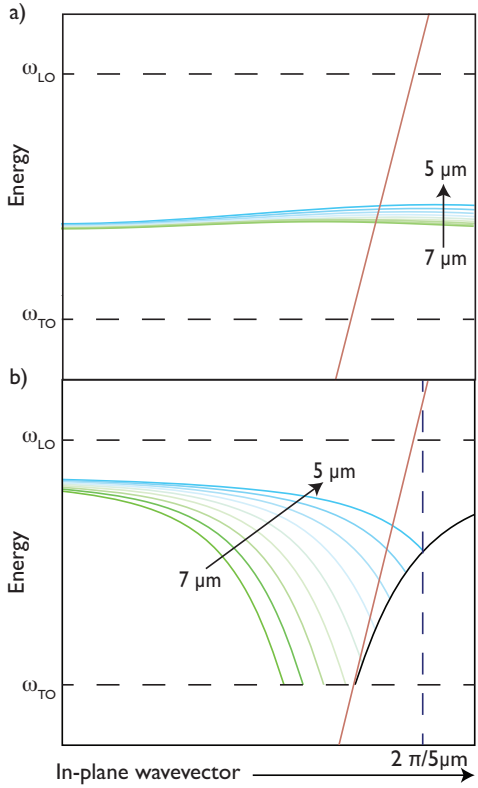}
\centering
\caption{\label{fig:FigSPhPFold} a) The tight binding dispersion of the monopolar mode is shown for a range of array periods from $5\mu$m to $7\mu$m. b) The dispersion of the first folded surface phonon polariton branch at a SiC/vacuum interface is illustrated by the coloured lines for array pitches between $5$ and $7\mu$m. The edge of the first Brillouin zone is illustrated by the dashed vertical line for period $5\mu$m, while the horizontal dashed lines show the borders of the Reststrahlen band. }
\end{figure}

When the substrate is periodically patterned a more in-depth analysis is needed as the normal modes of the system will now be given by Bloch waves delocalised over the array. On one hand this can lead to dispersive features in the dispersion of the localised phonon polaritons, due to dipolar coupling between the different cylinders. Such an effect can be captured by a tight binding model, that we already described in Ref. \onlinecite{Gubbin16}. This procedure, whose details can be found in Appendix C, lead to frequencies for the monopolar and transverse dipolar modes dependent upon the in-plane wavevector. While this effect is important for the monopolar mode, as shown in Fig.~\ref{fig:FigSPhPFold}a, it is negligible for the transverse modes. This can be understood noticing that, as shown in Ref. \onlinecite{Caldwell13}, the charge imbalance of the transverse modes is localised in the cylinders whereas the monopolar mode induces a charge imbalance between the cylinders and the the substrate, thus efficiently coupling the cylinders between them. In the following we will assume that all transverse modes are dispersionless for the array periods considered.

On the other hand the periodicity of the array causes the dispersion of the surface modes bound to the substrate to be folded back into the first Brillouin zone of the lattice, thus existing at experimentally accessible wavevectors within the lightcone. This folding as a function of array periodicity is illustrated in Fig.~\ref{fig:FigSPhPFold}b for square arrays of varying period. Such a tuneability can bring the localised and surface modes in resonance, and their coupling can not be reduced to a simple shift as in the previous Section, but it becomes necessary to consider the hybridization of the different bare modes.
In order to do this we use an extension of the Hopfield theory we used in Ref. \onlinecite{Gubbin16}. Notice that, while we recently also developed an extension of the Hopfield theory to inhomogeneous, lossy media \cite{Gubbin16b}, capable to give a description of the resonances without adjustable parameters, and including losses in a more consistent and fundamental way, here we prefer to rely on numerical simulations to fit the coupling parameters, and to use real frequencies instead that complex ones, calculating linewidths only in a second step, as this method is more transparent and readily applicable to the design and optimisation of resonator samples. Our approximate results will then be compared with numerical simulations performed using the quasi-normal mode theory described in Appendix A.

When only a single branch of the folded SPhP lies in the neighbourhood of the resonator lower laying modes, the system, neglecting losses, may be described by an Hamiltonian composed of two terms. The first 
\begin{align}
\mathcal{H}_{0} = \mathrm{\hbar \sum\limits_{k_{\parallel}}} \biggr[& \mathrm{\omega_{k_{\parallel}}^{m} \hat{a}_{k_{\parallel}}^{\dagger}\hat{a}_{k_{\parallel}} + \omega^{t1} \hat{b}_{k_{\parallel}}^{\dagger}\hat{b}_{k_{\parallel}}} \\&\mathrm{+  \omega^{t2} \hat{c}_{k_{\parallel}}^{\dagger}\hat{c}_{k_{\parallel}} + \omega_{k_{\parallel}}^{s} \hat{d}_{k_{\parallel}}^{\dagger}\hat{d}_{k_{\parallel}}}\biggr],\nonumber
\end{align}
describes the uncoupled modes, where $\mathrm{k}_{\parallel}$ is the in-plane wavevector,
$\omega_{\mathrm{k}_{\parallel}}^{\mathrm{m}}$ is the real part of the dispersive frequency of the monopolar mode obtained by the tight-binding approach described in Appendix C, 
$\omega^{\mathrm{t}1}$ and $\omega^{\mathrm{t}2}$ the real parts of the frequencies of the two lowest transverse $m=1$ modes, $\omega_{\mathrm{k}_{\parallel}}^{\mathrm{s}}$ the real part of the  surface mode from \Eq{surfacedispersion},  and  $\hat{\mathrm{a}}_{\mathrm{k}_{\parallel}}$,$\hat{\mathrm{b}}_{\mathrm{k}_{\parallel}}$,$\hat{\mathrm{c}}_{\mathrm{k}_{\parallel}}$,$\hat{\mathrm{d}}_{\mathrm{k}_{\parallel}}$ are the respective annihilation operators obeying bosonic commutation rules.
The second term describes instead the coupling between surface and resonator modes
\begin{align}
\mathrm{\mathcal{H}_{int}}   =\mathrm{ \hbar\sum\limits_{k_{\parallel}}} \biggr[& \mathrm{f_{0}\left(\hat{a}_{k_{\parallel}}^{\dagger}\hat{d}_{k_{\parallel}} + \hat{d}_{k_{\parallel}}^{\dagger}\hat{a}_{k_{\parallel}}\right)}\\
&+\mathrm{ g_{0}\left(\hat{b}_{k_{\parallel}}^{\dagger}\hat{d}_{k_{\parallel}} + \hat{d}_{k_{\parallel}}^{\dagger}\hat{b}_{k_{\parallel}}\right)}\nonumber \\
& + \mathrm{ h_{0}\left(\hat{c}_{k_{\parallel}}^{\dagger}\hat{d}_{k_{\parallel}} + \hat{d}_{k_{\parallel}}^{\dagger}\hat{c}_{k_{\parallel}}\right)}\biggr],\nonumber
\end{align}
where $\mathrm{f}_0,\;\mathrm{g}_0$ and $\mathrm{h}_0$ are the coupling rates of the monopole and lowest two transverse modes respectively with the quasi-resonant surface branch. The rotating wave approximation has been used, as the condition $\mathrm{f_0,\; g_0,\;h_0 \ll \omega_{k_{\parallel}}^{m}, \; \omega^{t1},\;\omega^{t2}, \; \omega_{k_{\parallel}}^{s}}$ is safely satisfied, with a coupling to frequency ratio of the order of $10^{-2}$ \cite{Gubbin16}.  
\begin{figure}
	\includegraphics[width=0.5\textwidth]{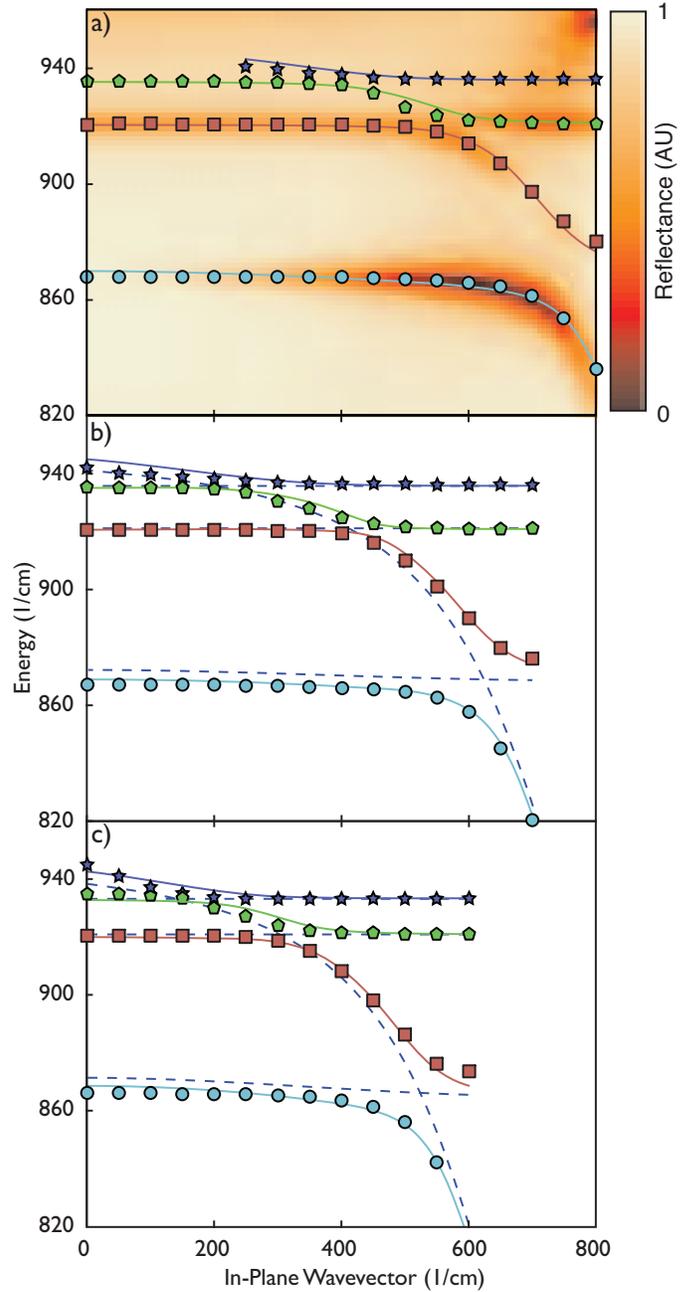}	
	\caption{\label{fig:Figure4} Resonances of the pillar array coupled to the substrate as a function of the in-plane wavevector for array periods a) $6\mu$m, b) $6.5 \mu$m and c) $7 \mu$m. The symbols represent the real parts of the frequencies of the four coupled eigenmodes calculated using quasi-mode theory as from Appendix A. The solid lines are instead the fits obtained solving Eq.~\ref{eq:Charac2}.
Each plot is truncated to restrict to the first Brillouin zone. Dotted lines represent the dispersions of the bare mode. The top panel is overlaid on a reflectance map calculated by finite element simulations to demonstrate the accuracy of the numerical methods employed.}
\end{figure}
The normal modes of the coupled system $\mathcal{H}=\mathcal{H}_0 + \mathcal{H}_{\mathrm{int}}$ can be expressed in the form of linear superpositions of the bare modes 
\begin{equation}
\mathrm{\hat{Y}_{k_{\parallel}}^i = m_{k_{\parallel}}^i \hat{a}_{k_{\parallel}} + n_{k_{\parallel}}^i \hat{b}_{k_{\parallel}} + o_{k_{\parallel}}^i \hat{c}_{k_{\parallel}} + p_{k_{\parallel}}^i \hat{d}_{k_{\parallel}}},
\end{equation}
where the Hopfield coefficients $\mathrm{m_{k_{\parallel}}^i, \;n_{k_{\parallel}}^i, \;o_{k_{\parallel}}^i \; \; p_{k_{\parallel}}^i}$ can be found 
solving the eigenproblem
\begin{equation}
\left[ \mathcal{M}_\mathrm{{k_{\parallel}}}-\mathrm{\omega_{k_{\parallel}}^i} \right]
\left( \begin{array}{c}
\mathrm{m_{k_{\parallel}}^i }\\
\mathrm{n_{k_{\parallel}}^i}  \\
\mathrm{o_{k_{\parallel}} ^i }\\
\mathrm{p_{k_{\parallel}}^i}  \end{array} \right)
= 0,
\label{eq:Charac2}
\end{equation}
where $\mathcal{M}_\mathrm{{k_{\parallel}}}$ is the Hopfield matrix
\begin{equation}
\mathcal{M}_\mathrm{{k_{\parallel}}}=
\left( \begin{array}{cccc}
\mathrm{\omega_{k_{\parallel}}^{m}}  & 0 & 0 & \mathrm{f}_0 \\
0 & \mathrm{\omega^{t1} }  & 0 & \mathrm{g}_0 \\
0 & 0 & \mathrm{\omega^{t2} }  & \mathrm{h}_0 \\
\mathrm{f}_0 & \mathrm{g}_0 & \mathrm{h}_0 & \mathrm{\omega_{k_{\parallel}}^{s} }\end{array} \right),
\label{eq:Charac}
\end{equation}
and the eigenvalue $\mathrm{\omega_{k_{\parallel}}^i}$ is here to be interpreted as the real part of the $\mathrm{i^{\text{th}}}$ quasi-normal mode frequency $\mathrm{\tilde{\omega}_{k_{\parallel}}^i}$. Results for array periods $6 \mu$m, $6.5\mu$m and $7\mu$m, are shown in Fig.~\ref{fig:Figure4}a, b, and c, where the symbols represent the real part of the complex quasi-mode frequencies obtained through the numerical procedure described in Appendix A, and the solid lines are obtained by solving \Eq{eq:Charac2} and fitting for $\mathrm{\omega^{t1}, \omega^{t2}, f_0, g_0, h_0}$, and the $\alpha$ and $\zeta$ parameters described in Appendix C. The plots are truncated before the edge of the first Brillouin zone to avoid inclusion of additional SPhP branches. \\
Excellent agreement is achieved between the two approaches. The peak Rabi frequency calculated is $15.84/$cm, representing $\approx 2\%$ of the bare mode energy. The calculated frequencies for the highest energy polariton branch are systematically lower in the fitted data; this is due to coupling of the surface mode to higher energy, closely-spaced resonances near the asymptotic SPhP frequency which have been omitted for simplicity from the Hopfield diagonalization.
The Hopfield coefficients weighting the bare components of the four coupled modes are shown in Fig.~\ref{fig:Figure5}a-d for an array period of $6\mu$m, from which it is clear that a substantial hybridization between the different modes occurs.\\
\begin{figure}
	\includegraphics[width=0.4\textwidth]{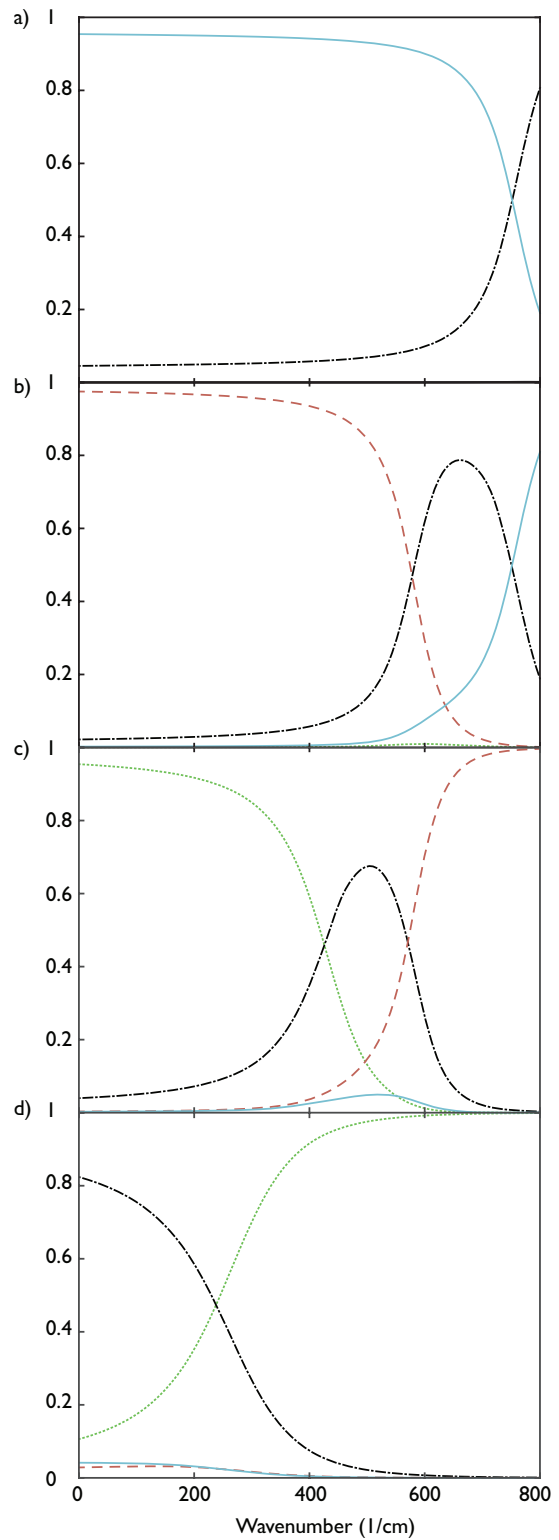}	
	\caption{\label{fig:Figure5} Plots of the absolute magnitude of the Hopfield coefficients for array period $6\mu$m. Panels a-d show the bare components of each coupled mode individually, in order of ascending energy. The solid line represents the monopole coefficient, the dashed line represents the dipole transverse, dotted the quadrupole transverse and the dot-dashed line represents the surface mode.}
\end{figure}

\begin{figure}
	\includegraphics[width=0.4\textwidth]{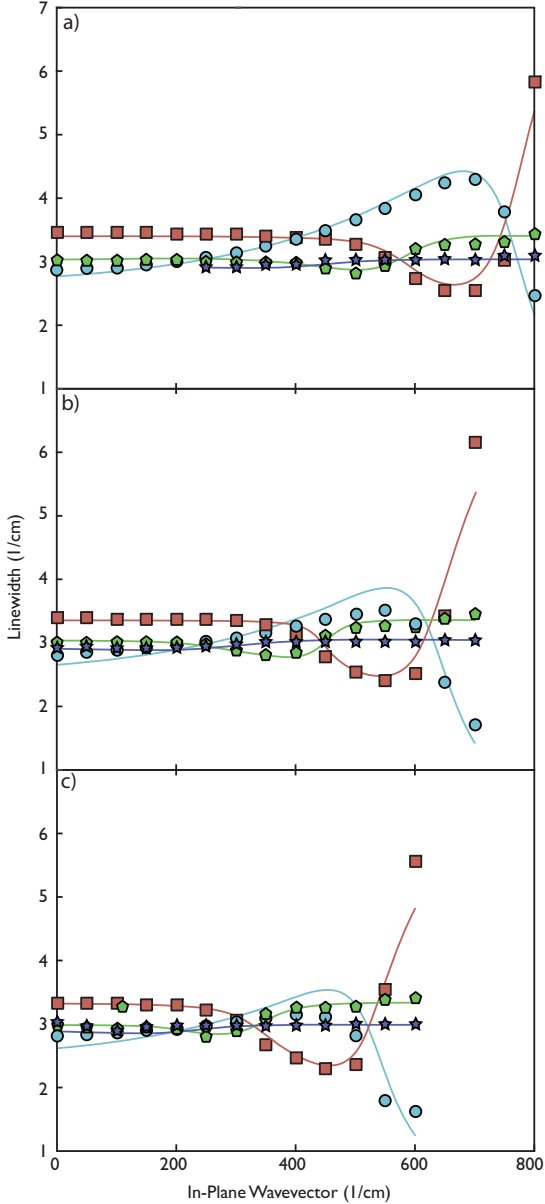}	
	\caption{\label{fig:Figure6} Normal mode linewidths for array periods a) $6\mu$m, b) $6.5\mu$m and c) $7\mu$m. Symbols indicate numerical values. Lines indicate theoretical fits calculated utilising Eq.~\ref{eq:LW}. Each plot is truncated to restrict to the first Brillouin zone.}
\end{figure}
As the polaritonic modes are linear superpositions of the bare modes, their linewidths are expected to vary predictably as sums of the linewidths of the bare modes weighted by the square of the Hopfield coefficients  \cite{Hopfield1958}. The linewidth of the $\mathrm{i^{\text{th}}}$ coupled mode can thus be
written in terms of the linewidths of the bare modes $\Gamma_{k_{\parallel}}^{\mathrm{m}},\Gamma_{k_{\parallel}}^{\mathrm{t1}},\Gamma_{k_{\parallel}}^{\mathrm{t2}}$, and $\Gamma_{k_{\parallel}}^{\mathrm{s}}$ and of the Hopfield coefficients as
\begin{equation}
\mathrm{\Gamma_{k_{\parallel}}^{i} = \lvert m_{k_{\parallel}}^i\rvert^2 \Gamma_{k_{\parallel}}^m + \lvert n_{k_{\parallel}}^i\rvert^2 \Gamma_{k_{\parallel}}^{t1} + \lvert o_{k_{\parallel}}^i\rvert^2 \Gamma_{k_{\parallel}}^{t2} + \lvert p_{k_{\parallel}}^i\rvert^2 \Gamma_{k_{\parallel}}^s},
\label{eq:LW}
\end{equation}
which can be fitted to the imaginary part of the modal frequency calculated using the approach detailed in Appendix A. 
Numerical results for array periods $6 \mu$m, $6.5\mu$m and $7\mu$m are given by the symbols in Fig.~\ref{fig:Figure6}. The dispersive surface mode linewidth is taken from \Eq{surfacedispersion} and the known dielectric function. Fits are carried out for the two transverse linewidths $\mathrm{\Gamma_{k_{\parallel}}^{t1}, \Gamma_{k_{\parallel}}^{t2}}$ and for the monopolar linewidth. The monopole linewidth is dispersive and fits to a phenomenological $\mathrm{a + b k_{\parallel}^2}$ relationship, verified by fitting to the dispersive monopole linewidth in the absence of the substrate. The results are given by the solid lines in Fig.~\ref{fig:Figure6}. Surprisingly good agreement between theory and numerics are achieved despite the simplicity of the model.\\ 

\section{Conclusion}
We have investigated the morphology and substrate dependant phonon polariton resonances of cylindrical SiC nano resonators by quasi-normal modal analysis. Starting from the resonances of a single, free standing cylinder, and then considering the impact of resonant coupling with surface phonon polariton modes sustained by the substrate, we were able to develop a complete and consistent understadning of those resonators. 
The present work will allow for the easy design of novel samples with bespoke resonances, and it shine light on the nature of the hybrid localised-surface resonances, that will permit further investigations to explore different geometry and sample materials.\\

\vspace{1cm}
\section{Acknowledgements}

S.A.M. acknowledges support from EPSRC programme grants EP/L024926/1 and EP/M013812/1, plus ONR Global, the Royal Society, and the Lee-Lucas Chair in Physics.
S.D.L. is Royal Society Research Fellow and he acknowledges support from EPSRC grant EP/M003183/1.

\appendix
\begin{figure*}[t]
\includegraphics[width=0.7\textwidth]{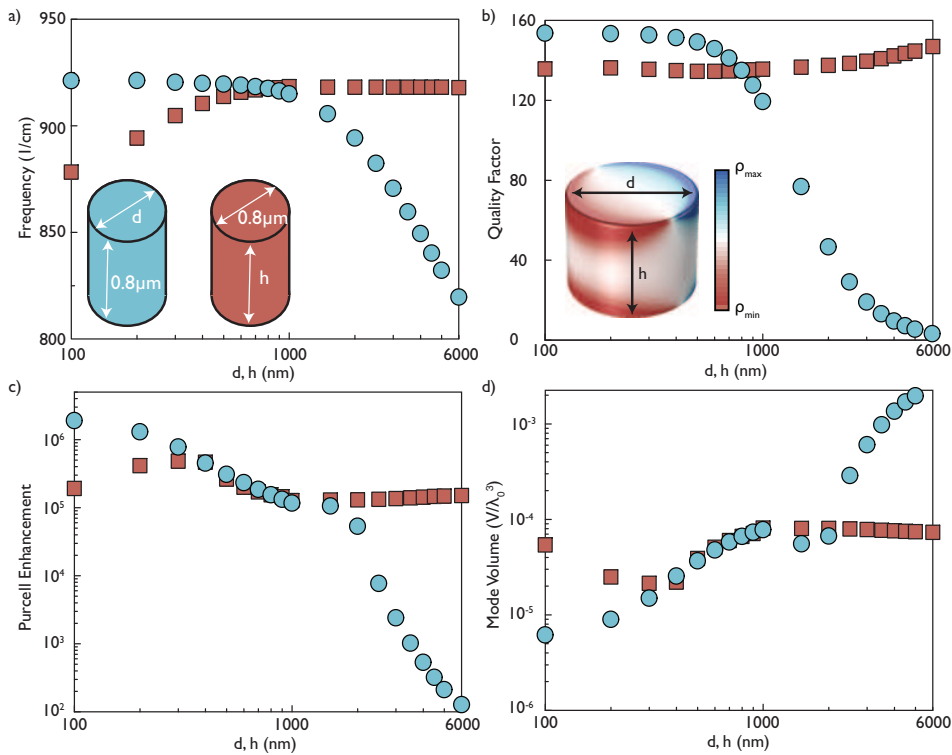}
\centering
\caption{\label{fig:Supp2}  
a) Real part of the resonant frequency of the dipolar transverse mode as a function of diameter at $\mathrm{h}=0.8\mu$m height (circles) and as a function of height at $\mathrm{d}=0.8\mu$m diameter (squares).
b) Quality factor of the dipolar transverse mode as a function of diameter at $0.8\mu$m height (circles) and as a function of height at $0.8\mu$m diameter (squares). Inset shows the surface charge distribution $\rho$ of the dipolar transverse mode for a cylinder of height $\mathrm{h}=0.8\mu$m and diameter $\mathrm{d}=1\mu$m. 
e)  Purcell enhancement of the dipolar transverse mode as a function of diameter at $0.8\mu$m height (circles) and as a function of height at $0.8\mu$m diameter (squares).
f) Mode volume of the dipolar transverse mode in units of the free space mode volume $\lambda_0^3$ as a function of diameter at $0.8\mu$m height (circles) and as a function of height at $0.8\mu$m diameter (squares).}
\end{figure*}
\section{Quasinormal Mode Theory}
Under fairly general assumptions \cite{Leung1994}, the electromagnetic fields radiated by an emitter in the resonator $\vec{\Upsilon}\left(\vec{r},\omega\right)=(\vec{E}\left(\vec{r},\omega\right), \vec{H}\left(\vec{r},\omega\right))$ can be linearly expanded onto a discrete set of quasi-normal modes $\tilde{\vec{\Upsilon}}_\mathrm{i}\left(\vec{r}\right)=(\tilde{\vec{E}}_\mathrm{i}\left(\vec{r}\right), \tilde{\vec{H}}_\mathrm{i}\left(\vec{r}\right))$
\begin{equation}
\mathrm{\vec{\Upsilon}\left(\vec{r},\omega\right) = \sum_{i} \alpha_{i}\left(\omega\right) \tilde{\vec{\Upsilon}}_i\left(\vec{r}\right)},
\end{equation}
where $\alpha_\mathrm{i}\left(\omega\right)$ is a complex coefficient describing the relative contribution of the $\mathrm{i}^{\text{th}}$ mode, and it has a pole at the complex modal frequency $\tilde{\omega}_\mathrm{i}$. The quasi-normal modes of the system are found using an iterative procedure to fit to a Pade approximated pole-like response function in the complex frequency plane \cite{Bai2013}. The iterative procedure is carried out utilising the COMSOL Multiphysics FEM solver iteratively driven by a MATLAB code.\\
On calculating the complex modal frequencies of the system we can immediately calculate the quality factor 
\begin{equation}
	\mathrm{Q}_\mathrm{i} = \frac{\mathrm{Re}\left[\tilde{\omega}_\mathrm{i}\right]}{2\mathrm{Im}\left[\tilde{\omega}_\mathrm{i}\right]},
\label{eq:QF}
\end{equation} 
as well as any other quantity of interest. In the neighbourhood of the complex frequency $\tilde{\omega}_\mathrm{i}$ it is an excellent approximation that the field scattered by the resonator is linearly proportional to the field of the quasinormal mode. It is therefore possible to calculate the mode volume through the equation
\begin{equation}
\mathrm{V_{i} = \frac{\int \left[\tilde{\vec{E}}_i \cdot \frac{\partial\left(\omega\vec{\epsilon}\left(\vec{r},\omega\right)\right)}{\partial \omega} \tilde{\vec{E}}_i - \tilde{\vec{H}}_i \cdot \frac{\partial\left(\omega\vec{\mu}\left(\vec{r},\omega\right)\right)}{\partial \omega} \tilde{\vec{H}}_i\right]\mathrm{d}^3 \vec{r}}{2 \epsilon_0 n\left(\vec{r}_{\mathrm{max}}\right)^2 \left[\tilde{\vec{E}}_i\left(\vec{r}_{\mathrm{max}}\right)\cdot \vec{u}\right]^2}},
\label{eq:ModeVolCorrect}
\end{equation}
where the proportionality constants cancel from the numerator and denominator and the fields are evaluated very close to the complex resonant frequency \cite{Sauvan2013}. The integral over all space is regularised utilising a perfectly matched layer. At the complex resonant frequency both terms in the numerator are divergent, the sum however is not.\\
The dispersive dielectric function of the matter is defined by the fitting parameters of Pitman \cite{Pitman2008} for 3C-SiC to the functional form 
\begin{equation}
	\epsilon\left(\omega\right) = \epsilon_{\infty} + \frac{\omega_{\mathrm{LO}}^2 \left(\epsilon_0-\epsilon_{\infty}\right)}{\omega_{\mathrm{LO}}^2 - \frac{\epsilon_0}{\epsilon_{\infty}} \omega^2 - i \frac{\epsilon_0}{\epsilon_{\infty}} \gamma \omega}.
	\label{eq:DieFun}
\end{equation}
This was used over an interpolated dielectric function to allow for analytic continuation to the complex frequency plane. In passing from the standard definition of the mode volume to Eq.~\ref{eq:ModeVolCorrect} material dispersion is accounted for by taking the derivatives of the system dielectric function and the permeability. These derivatives are especially important in polar dielectric systems where the inflection of the dielectric function occurs entirely over the narrow bandwidth of the Reststrahlen band. The ratio of of the dielectric function $\epsilon\left(\omega\right)$ in Eq.~\ref{eq:DieFun} and the derivative $\omega \partial \epsilon\left(\omega\right)/ \partial \omega$ exceeds unity throughout the Reststrahlen band, often lying between 10-100. This means that the contribution from the electric field energy in the polar dielectric increases by 1-2 orders of magnitudes resulting in a substantial decrease in the achievable field confinements compared to the rudimentary Eq.~\ref{eq:ModeVolStandard}, sometimes falling by up to two orders of magnitude. Physically this result arises from energy lying in the potential energy of the oscillating ions rather than in the electric field as illustrated in Fig.1b \cite{Khurgin15}.\\
Finally the Purcell enhancement for the $\text{i}^{\text{th}}$ mode may be calculated from the usual equation
\begin{equation}
	\mathrm{F_{P}} = \frac{\Gamma}{\Gamma_0} = \frac{3}{4 \pi^2} \left(\frac{\lambda_0}{\mathrm{n}}\right)^3 \mathrm{Re}\left[\frac{\mathrm{Q_i}}{\mathrm{V_i}}\right],
	\label{eq:AnaPur}
\end{equation}
where $\Gamma \left(\Gamma_0\right)$ are the decay rates of a dipole in the presence of the resonator (in free space), $\mathrm{n}$ is the refractive index at the dipole location, $\lambda_0$ is the free space wavelength, and $\mathrm{Q_i}$ and $\mathrm{V_i}$ are as defined in the previous equations.\\

\section{The Transverse Dipolar Mode}
The resonant frequency of the dipolar transverse mode is investigated over the 2D parameter space in  Fig.~\ref{fig:Supp2}a. Squares correspond to a diameter scan at $\mathrm{h}=0.8\mu$m and circles to a height scan at $\mathrm{d}=0.8\mu$m. A red shift in the resonant frequency with increased diameter is observed, this occurs due to increased screening between charges at each cylinder edge. This increased screening pushes the mode frequency toward the transverse optical phonon frequency with accompanying drop in quality factor, shown in Fig.~\ref{fig:Supp2}b as observed for the monopolar mode in Fig.~\ref{fig:Figure1}b. In the large height limit $\mathrm{h} \geq 1\mu$m the resonant frequency reaches an asymptote as the dipoles at the end facets decouple, scanning the height weakly affects the quality factor as the mode is strongly localised at the cylinder end facets.\\
The surface charge distribution is illustrated on the inset in Fig.~\ref{fig:Supp2}a. 
The Purcell enhancement of the transverse mode is plotted  in Fig.~\ref{fig:Supp2}c. Strong increases are observed in smaller geometries, exceeding $10^6$ as $\mathrm{d} \to 0.1\mu$m. These Purcell enhancements correspond to ultra-small mode volumes less than $10^{-4} \lambda_0^3$ as shown in Fig.~\ref{fig:Supp2}d.\\

\begin{figure}
\includegraphics[width=0.45\textwidth]{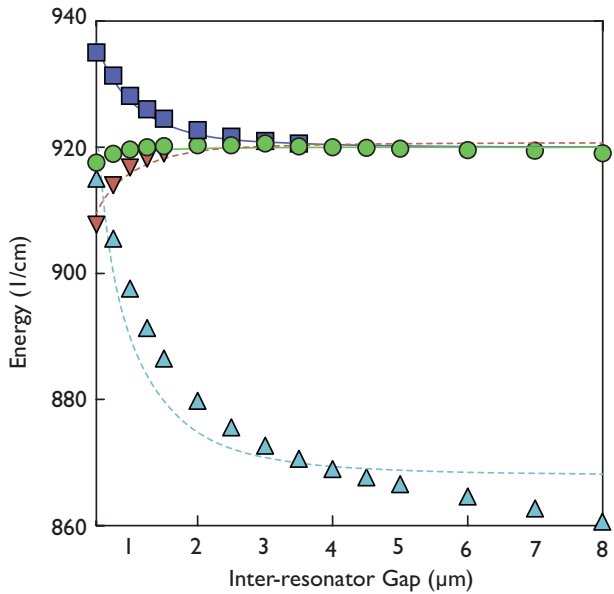}
\centering
\caption{\label{fig:Supp4}   Symbols represent the real resonant frequencies calculated from the pole fitting algorithm at the $\mathrm{k}_{\parallel} = 0$ point for a square array of cylinders of height $0.8\mu$m and diameter $1\mu$m. Purple squares (green circles) represent the longitudinal (lowest transverse) mode in free cylinders. Blue triangles (red inverted triangles) represent the longitudinal (lowest transverse) mode for cylinders in contact with an SiC substrate. The overlapping lines represent a fit with a simple dipole-dipole coupling model.}
\end{figure}
\section{Periodic Resonator Arrays}
To achieve experimentally measurable observables it is necessary to measure arrays of resonators. In this paper we restrict to square arrays of resonators. The modes of the coupled array is taken as a linear combination of the individual resonators quasi-normal modes $\tilde{\vec{E}}_{\mathrm{i}}\left(\vec{r}\right)$ along a straight line parallel to the illumination wavevector which is taken parallel to an array principal axis. The dispersion of the mode of the periodic system may be approximated by the solution calculated in Ref. \onlinecite{Yariv1999} for lossless systems in the tight binding approximation as 
\begin{equation}
\mathrm{\omega_{k_{\parallel}}^{\mathrm{i}} = \omega^{\mathrm{i}}\left(1 - \frac{\Delta \gamma}{2} + \left(\beta_1-\gamma_1\right) \cos\left(k_{\parallel} R\right)\right)},
\end{equation}
where $\omega_{\mathrm{k_{\parallel}}}^{\mathrm{i}}$ is the dispersive frequency, $\mathrm{k}_{\parallel}$ is the in-plane wavevector, $\omega^{\mathrm{i}}$ is the frequency of the isolated resonator mode, and the $\Delta \gamma, \; \beta_1, \; \gamma_1$ are as defined in Ref. \onlinecite{Yariv1999}.

The strength of inter resonator coupling is investigated for square arrays of cylinders in vacuum and on a substrate. The cylinders are of height $0.8\mu$m and diameter $1\mu$m. The inter-resonator gap is varied to assess the coupling at the $\mathrm{k}_{\parallel} = 0$ point. Symbols in Fig.~\ref{fig:Supp4} represent the real frequencies calculated from the pole fitting algorithm, lines represent fits to the results assuming a simple $1/\mathrm{r}^3$ dipole-dipole coupling. In each case the longitudinal mode and lowest lying transverse mode are studied. The longitudinal mode blue shifts as the inter-resonator gap is decreased, this is because the dipoles are orientated in the same direction along the cylinder long axis, repulsing each other. Conversely the transverse mode red shifts as the inter-resonator gap is decreased, because dipoles facing each other on neighbouring cylinders are aligned in opposite directions and they attract each other. Larger shifts are observed for the resonators on substrate, this is because the substrate is highly reflective in the Reststrahlen band and more radiative emission from each resonator propagates to the next. From Fig. \ref{fig:Supp4} it is also clear that the monopolar mode is much more dispersive than the transverse one. This is a general feature due to the fact that the monopolar mode generates a flow of charge between the pillars and the inter-pillar surfaces \cite{Caldwell13}, increasing the coupling.
For this reason, in the fitting procedure outlined in the text, we only considered the dispersion of the monopolar mode, leading to the two fitting parameters $\alpha = \mathrm{\omega^m\left(1 - \Delta\gamma/2\right)}$ and $\zeta =\mathrm{ \omega^m \left(\beta_1 - \gamma_1\right)}$, with $\omega^{\mathrm{m}}$ the frequency of the monopolar mode in the single cylinder.\\

\end{document}